# Magnetism and Superconductivity of S-substituted FeTe

A. Ciechan[a,*], M.J. Winiarski[b], M. Samsel-Czekała[b]

[a]*Institute of Physics, Polish Academy of Sciences, al. Lotników 32/46, 02-668 Warsaw, Poland*
[b]*Institute of Low Temperature and Structure Research, Polish Academy of Sciences, ul. Okólna 2, 50-422 Wrocław, Poland*

**Abstract**

The influence of a partial substitution with sulphur into Te sites on the crystal, electronic and magnetic structures of FeTe is investigated by DFT calculations. The results reveal a phase transition from the antiferromagnetic double-stripe order for pure FeTe to the single-stripe order for S-substituted samples, which coincides with the previously observed appearance of the superconducting state. The magnetic transition is caused by the variations of the average chalcogen position in the unit cell. The analyzed normal-state properties of Fe(Te,S) and Fe(Se;S) compounds allow a detection of the well resolved nesting-driven magnetic fluctuations only for superconducting samples, consistent with their antiferromagnetic ground state. Thus, the role of an S-substitution is a suppression of the double-stripe antiferromagnetic order to give rise to the single-stripe correlations, which are associated with an occurrence of superconductivity in Fe(Te,S) solid solutions.

(The figures in this article are in colour in the online version)

*Keywords:* high-Tc superconductors, magnetically ordered materials, metals and alloys, crystal structure, electronic band structure, electronic properties, phase transitions

## 1. Introduction

Superconductivity in iron chalcogenides occurs in the close proximity to magnetic instabilities and, hence, seems to be mediated by spin fluctuations. Fe(Se,Te) solid solutions are superconducting with the maximal critical temperature ($T_c$) reaching 15 K for FeTe$_{0.5}$Se$_{0.5}$ at ambient pressure and 37 K for pure FeSe under hydrostatic pressure [1, 2, 3, 4, 5, 6]. FeTe exhibits an antiferromagnetic (AFM) order at low temperature [1, 7, 8], while the superconducting state does not appear even under such high pressure as 19 GPa [9, 10]. Instead, the pressure of 2 GPa applied to pure FeTe induces the magnetic phase transition from an AFM to ferromagnetic (FM) state [11]. Interestingly, superconductivity emerges in FeTe under Se or S substitutions for Te sites in bulk crystals [1, 2, 12, 13, 14, 15, 16] and under tensile stress in thin epitaxial films [17]. Also FeS crystallizes in a similar tetragonal structure [18], however, the compound is neither superconducting nor magnetic.

In FeTe$_{1-x}$S$_x$, the superconducting state appears for x = 0.03 and $T_c$ reaches the value of 10 K [12, 13, 14, 15, 16]. Unlike the Fe(Te,Se) series, there exists the solubility limit of 30% S content in Fe(Te,S) caused by a large difference between S and Te ionic radii [16]. Previously, it was indicated that the limit is 15% S or even less [14, 15], which is probably mainly the result of a difference between nominal and final compositions of studied samples. It seems that both Se and S substitutions suppress the magnetic order of FeTe thereby leading to the superconducting state. It implicates also a stabilization of the tetragonal crystal structure of FeTe and a decrease of the distance between chalcogen atoms and the Fe plane. Negative biaxial pressure has a similar effect on the crystal structure, which induces superconductivity in FeTe thin films grown on SrTiO$_3$ or MgO substrate [17].

All iron chalcogenides reveal a similar band structure, according to both photoemission spectra [19, 20, 21, 22, 23, 24] and DFT calculations [25, 26, 27, 28, 29]. A few hole and electron bands are crossing the Fermi level ($E_F$) near the Γ and M points, respectively. It is believed that the nesting of the hole-like and electronlike Fermi surface sheets induces spin fluctuations and leads to superconducting pairing. Such fluctuations characterized by the wave vector of ($\pi, \pi$) were observed for Fe(Se,Te) superconductors below $T_c$ in neutron scattering and nuclear magnetic resonance measurements [30, 31]. However, these fluctuations correspond to a single-stripe AFM order, while FeTe exhibits a double-stripe AFM order with the ($\pi, 0$) wave vector [8, 32]. It seems that magnetic correlations strongly evolve in Fe(Te,Se) solid solutions and superconducting state occurs when the ($\pi, 0$) fluctuations are sufficiently suppressed and the ($\pi, \pi$) fluctuations become dominant [33, 34, 35].

The effects of hydrostatic and non-hydrostatic pressure on the magnetic and electronic structures of FeTe have been investigated by first principle calculations [36, 37]. The results indicate that the magnetic ordering and superconductivity in iron chalcogenides is closely connected with the crystal structure. Interestingly, the tensile strain applied to FeTe enhances the ($\pi, \pi$) nesting [38], analogously to the case of superconducting Fe(Te,Se) systems [39, 40].

In this work, we study the influence of sulfur on the mag-netic structure of Fe(Te,S). We obtain the transition from the double-stripe to the single-stripe order for sulfur contents higher than 2%, which is related to the variation of the averaged chalcogen position in the unit cell. This finding indicates that superconductivity in the studied systems is strongly related to the AFM fluctuations with the $(\pi,\pi)$ nesting vector. An analysis of the Fermi surface and spin susceptibilities obtained from the band structure of some iron chalcogenides suggests a presence of the well resolved nesting-driven magnetic fluctuations only for superconducting samples. This further confirms correlation between the $(\pi,\pi)$ spin fluctuations and the appearance of the superconducting state.

*Corresponding author. Tel.: +48 22 1163 191; fax: +48 22 8430 926.
Email address: ciechan@ifpan.edu.pl (A. Ciechan)

## 2. Computational methods

The examination of the effect of an S-substitution into Te sites in FeTe$_{1-x}$S$_x$ systems was performed within the density functional theory (DFT) in the generalized-gradient approximation (GGA) of the exchange-correlation potential [41, 42]. We used the pseudopotential method, based on plane-waves and Projector-Augmented Waves, implemented in the QUANTUM ESPRESSO code [43].

The choice of the adequate exchange-correlation functional (the local-density approximation, LDA, or GGA) used to obtain the proper characteristics of the system was widely discussed for iron pnictides [44]. It was noticed that the GGA generally overestimates magnetic stabilization energies and the local magnetic moment of Fe atoms. On the other hand, the GGA better reproduces the crystal structure and magnetic tendencies of analysed systems, which is extremely useful in our considerations of a magnetic ground state. Moreover, iron chalcogenides differ from pnictides in many respects, including their chemistry and magnetic ground state [29]. Especially, the LSDA does not reproduce the AFM ordering observed in FeTe (for experimental lattice parameters) [28], which further makes the GGA method more appropriate than LDA within spin-polarized calculations. The non-magnetic band structure, in turn, is expected to be less sensitive to the choice of the exchange-correlation functional, similar to pnictides [44]. It should be mentioned, however, that the appropriate approximation of exchange-correlation functional depends usually on the considered properties. It was shown that the use of norm-conserving pseudopotentials in the standard LDA approach or inclusion of van der Waals interaction leads to reasonable results of structural parameters for iron chalcogenides within non-spin-polarized calculations [45, 46]. Furthermore, the employment of the full-relativistic pseudopotential with the spin-orbit coupling instead of our scalar relativistic PAW would further stabilizes the magnetic structure [47] and might shift the transition between two AFM states of Fe(Te,S). As far as we treat the computational results in qualitative rather than a quantitative way, the obtained changes of the crystal, electronic and magnetic structures induced by the S substitution into Te sites well converge with the emergence of superconducting state from antiferromagnetic state.

In used pseudopotential method, the kinetic energy cutoff of 40 Ry for wavefunctions and 240 Ry for charge density were employed, respectively. The calculations were performed with the $12 \times 12 \times 6$ **k**-point grid for the Brillouin zone (BZ) related to the $1 \times 1 \times 1$ unit cell with the elongated $c$-axis, containing 2 Fe atoms. It corresponds to the $6 \times 6 \times 3$ and $6 \times 6 \times 6$ **k**-point grids for the folded Brillouin zones (FBZ) of the $2 \times 2 \times 2$ and $2 \times 2 \times 1$ supercells, respectively. However, the Fermi surface cuts (and related band properties) were calculated with much denser grids of $201 \times 201$ **k**-points in $k_z$ planes for FBZ's of the $2 \times 2 \times 1$ supercells.

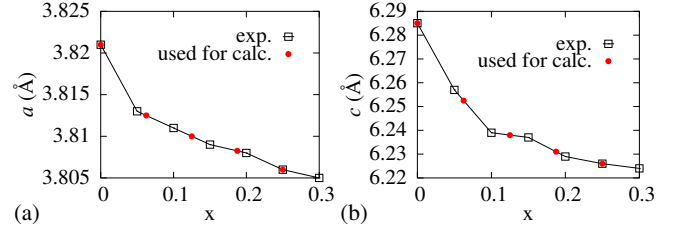

Figure 1: (a,b) The lattice parameters $a$ and $c$ taken from experiment [16] together with those obtained by interpolation and used in further calculations.

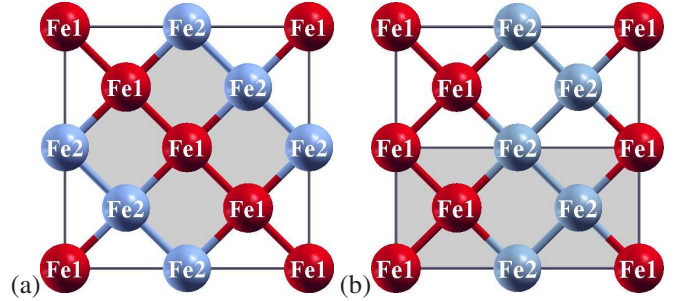

Figure 2: Schematic spin arrangements in $2 \times 2$ square plane of Fe atoms in antiferromagnetic phases of (a) single stripes (AFM1) and (b) double stripes (AFM2). The Fe atoms with different spin directions are indicated as Fe1 and Fe2. The magnetic unit cell in each case is marked by the shaded area.

The $2 \times 2 \times 2$ supercells were used to simulate magnetic structures of FeTe$_{1-x}$S$_x$ solid solutions with x = 0, 0.0625, 0.125, 0.1875 and 0.25. It allowed the study of the superconducting regime of 3-30% S [12, 13, 14, 15, 16]. We assumed the experimental lattice parameters $a$ and $c$ of the tetragonal phase of the PbO-type (space group P4/nmm) [16] by an interpolation for the considered concentrations (see Fig. 1). The tetragonal structure of Fe(Te,S) is well established experimentally [14]. We omitted in our considerations the monoclinic distortion of pure FeTe at low temperature because it leads to negligible changes of the magnetic and electronic properties [37, 48].

In the tetragonal structure of FeTe, iron atoms form a square lattice, while tellurium ions are located in the distance $h_{\text{Te}}$ from the Fe plane. Sulfur atoms, which are substituted into Te sites, change the local atomic structure and S atoms have a different distance $h_{\text{S}}$ from the Fe plane. We performed optimization of atomic positions $h_{\text{Te}}$ and $h_{\text{S}}$ of non-magnetic (NM) and magnetic phases in the $2 \times 2 \times 2$ supercells, which was carried out with a $10^{-3}$ Ry/Bohr convergence criterion on forces. We tested non-equivalent positions of S atoms in the supercell to obtain the averaged properties of the Fe(Te,S) systems.

Table 1: The calculated tellurium position, $h_{Te}$, of pure FeTe and its deviation from the experimental value of 1.77 Å, $\Delta h_{Te}$ (both in Å), magnetic stabilization energy, $|\Delta E|$ (in meV/f.u) for AFM1 and AFM2 state and the corresponding magnetic moment of Fe atoms, $\mu$ (in $\mu_B$/1 Fe at.).

|      | $h_{Te}$ | $\Delta h_{Te}$ | $|\Delta E|$ | $\mu$ |
|------|------|------|-----|------|
| AFM1 | 1.72 | 0.05 | 217 | 2.51 |
| AFM2 | 1.76 | 0.01 | 220 | 2.58 |
| NM   | 1.61 | 0.16 | 0   | -    |

In spin-polarized calculations, we took into account the single-stripe (AFM1) and double-stripe (AFM2) antiferromagnetic arrangements (Fig. 2) as these are the ground state of superconducting FeSe and non-superconducting FeTe samples, respectively [48, 49]. The antiparallel alignment of the Fe spin moments along the $c$ axis does not change total energy values significantly [37, 48] and we tested here only in-plane spin arrangements. We checked also other possible magnetic orders suggested in the literature [50], like FM or checkerboard AFM orders, but both turned out to be irrelevant to our considerations.

The electronic structure calculations for paramagnetic phases were performed employing $2\times2\times1$ supercells of pure FeTe and superconducting FeTe$_{0.875}$S$_{0.125}$ compositions. Such a calculation scheme folds the band structure to a respectively smaller BZ and makes complex its comparison with the ARPES measurements. However, the supercell approach appears to be convenient for studying the nesting nature, because now all the Fermi surface sheets are centered around the $\Gamma-Z$ line. Thus, the area of overlapping contours in the FBZ of the $2\times2\times1$ supercells reflects an intensity of the ideal nesting, which occurs with the $(\pi,\pi)$ vector in the BZ of the $1\times1\times1$ unit cell (see Ref. [51] for details).

The Lindhard spin susceptibility, derived from the band structure $\varepsilon_{m,\mathbf{k}}$, is calculated numerically from the formula:

$$\chi(\mathbf{q},\omega) = -\frac{1}{(2\pi)^3}\sum_{m,n}\sum_{\mathbf{k}\in \text{FBZ}}\frac{f(\varepsilon_{m,\mathbf{k}})-f(\varepsilon_{n,\mathbf{k+q}})}{\varepsilon_{m,\mathbf{k}}-\varepsilon_{n,\mathbf{k+q}}-\omega-i\delta},$$

where $f(\varepsilon_{m,\mathbf{k}}) = 0$ or 1 denotes the Fermi-Dirac distribution function at zero temperature. Thus defined spin susceptibility allows an estimation of spin fluctuations present in the system. This response function at the $\Gamma(0,0,0)$ point of FBZ reflects mainly the tendency to the nesting driven AFM1 ordering (which occurs at the $M_0(\pi,\pi)$ point of BZ) and a much weaker tendency to FM ordering (at the $\Gamma_0(0,0,0)$ point of BZ) [51, 52].

Electronic band properties of Fe(Te,S) are also compared with those of superconducting FeSe and non-superconducting FeS.

## 3. Results and discussion

### 3.1. Spin arrangements vs. crystal structure

The calculation results of the relaxed chalcogen position, $h_{Te}$, magnetic stabilization energy of AFM1 and AFM2 states with respect to the NM state, $\Delta E = E_{\text{AFM1/AFM2}} - E_{\text{NM}}$, and the corresponding magnetic moments of pure FeTe are collected in Table 1. As seen in the table, the obtained ground state of the system is AFM2, in agreement with the experimental findings [8, 32]. The calculated Te position, $h_{Te} = 1.76$ Å, agrees reasonably well with the experimental values of 1.76 – 1.77 Å [8, 9]. In turn, the structural optimization of the NM phase yields an underestimated $h_{Te}$ parameter. Interestingly, the proper crystal structure of superconducting iron chalcogenides and pnictides is obtained within spin-polarized calculations only [44, 53]. It suggests that the ground state of all these compounds is strongly magnetic. In the case of superconducting compounds, however, spin polarization leads to larger magnetic moments on Fe atoms than those observed in the experiment, which suggests a presence of spin fluctuations. Instead, for antiferromagnetically ordered FeTe, the Fe magnetic moment $\sim 2.6\mu_B$/Fe is in good agreement with the measured values of $2.25 - 2.54\mu_B$ [7, 8].

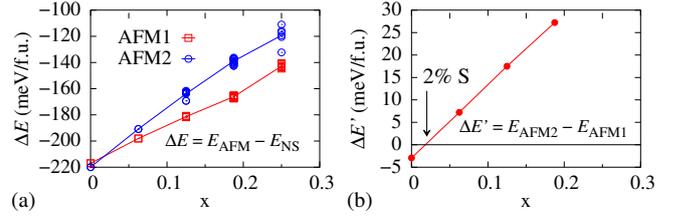

Figure 3: (a) Magnetic stabilization energy of antiferromagnetic (AFM1 and AFM2) states as a function of S content x. Multi-points indicate the results obtained for different S arrangements in the supercell, while lines plot the average energy of given magnetic phase. (b) The x dependence of the relative average total energy of the AFM2 with respect to the AFM1 state. The arrow indicates the critical content x = 0.02, for which AFM2 turns into the AFM1 ground state.

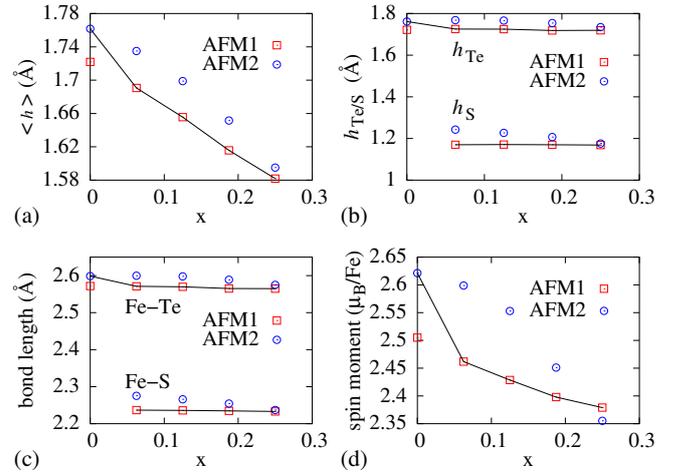

Figure 4: (a) The average chalcogen positions $<h>$ of all local $h_{Te}$ and $h_S$ positions in supercells as a function of S content x, where (b) $h_{Te/S}$ is the distance of the Te/S atom from the Fe plane. The corresponding x dependencies of (c) average bond lengths of Fe-Te and Fe-S atoms and (d) magnetic moment per one Fe atom.

For FeTe$_{1-x}$S$_x$ systems, the dependencies of the magnetic stabilization energy of both AFM1 and AFM2 states (with respect to the NM state) on a sulfur content x are shown in Fig-ure 3a. As seen in the figure, except for the case of the pure FeTe, the AFM1 state is found to be more stable than AFM2 for all x concentrations. Moreover, the arrangements of S atoms in the supercell turn out to be irrelevant to the obtained results, *i.e.* all energy values of AFM1 supercells are lower than those of AFM2 supercells for a fixed S content (see multi-points). By calculating relative values of the average total energy of AFM2 with respect to the AFM1 state, we estimated that the phase transition from the AFM2 to AFM1 state occurs for an S-content of about 2% (Fig. 3b).

Figure 4a shows the average chalcogen positions $<h>$ obtained from all Te and S sites in each supercells for given x of both AFM1 and AFM2 phases. The substitution of S atoms with smaller ionic radii than Te atoms decreases $<h>$ simultaneously with the compression of the lattice parameters $a$ and $c$ (Fig. 1). However, the local positions of both Te and S atoms are weakly dependent on the content x, being equal to $h_{Te} \approx 1.72/1.76$ Å and $h_S \approx 1.17/1.24$ Å for the AFM1/AFM2 state (Fig. 4b). In the NM phase, the positions are also nearly independent of the x concentration ($h_{Te} \approx 1.61$ Å and $h_S \approx 1.05$ Å), but significantly smaller. Similarly to the case of pure FeTe, only the values obtained for considered here magnetic phases are consistent with the experiment [14]. These values correspond to the Fe-Te bond length of $2.57 - 2.6$ Å and the Fe-S bond length of $2.24 - 2.28$ Å (Fig. 4c), while experimentally observed values amount to 2.61 Å and 2.26 Å, respectively. Interestingly, these bond lengths are also very close to those observed for both the pure FeTe and FeS [8, 18]. The situation resembles that of the Fe(Te,Se) solid solutions, in which Se and Te ions do not share the same atomic site, which leads to different $h_{Se}$ and $h_{Te}$ coordinates and lengths of the Fe-Te and Fe-Se bonds [54, 55]. Hence, the local symmetry of Fe(Te,S) is lower than the average P4/nmm, although the crystal structure remains tetragonal.

As seen in Fig. 4d, the local magnetic moment on the Fe atom is decreasing with the S content and generally is lower in the AFM1 state than in the AFM2 state. The trend is observed experimentally, although in real samples it coincides with the reduction of excess Fe atoms at interstitial sites of the Te layers [13]. The result is consistent also with DFT predictions for paramagnetic FeS, which exhibits a non-magnetic or weakly magnetic ground state of AFM1 type, dependent on used $h_S$ [25].

An alteration of a magnetic ground state was observed by the powder neutron diffraction in S-substituted FeTe [14]. It changes from commensurate ordering with the $(\pi, 0)$ propagation vector for FeTe to incommensurate one and finally the magnetic scattering is rapidly suppressed for samples with the S content $\sim 5\%$. Moreover, the resistivity and susceptibility measurements reveal a coexisting region of superconducting and antiferromagnetic states in the range of $x \in (0.03, 0.1)$ [13, 15]. Thus, the superconducting state appears in the system with the stable AFM1 phase. It is also worth noting that the differences between calculated total energy values of the AFM1 and AFM2 phases of Fe(Fe,S) are very small in the whole considered range of the S concentration. The pure FeTe is also on the verge of a magnetic instability and even small changes in the composition

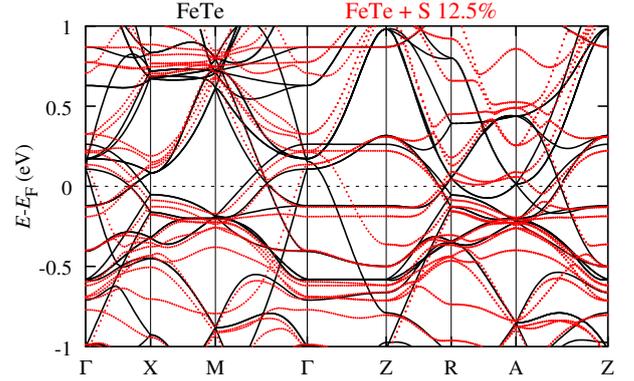

Figure 5: The non-magnetic band structure of FeTe and FeTe$_{0.875}$S$_{0.125}$. The positions $h_{Te}$ and $h_S$ are taken from relaxation of the spin-polarized states, *i.e.* AFM1 and AFM2, respectively. The results are shown for the $2 \times 2 \times 1$ supercell, corresponding to the folded Brillouin zone with its high symmetry points, denoted by Greek letters.

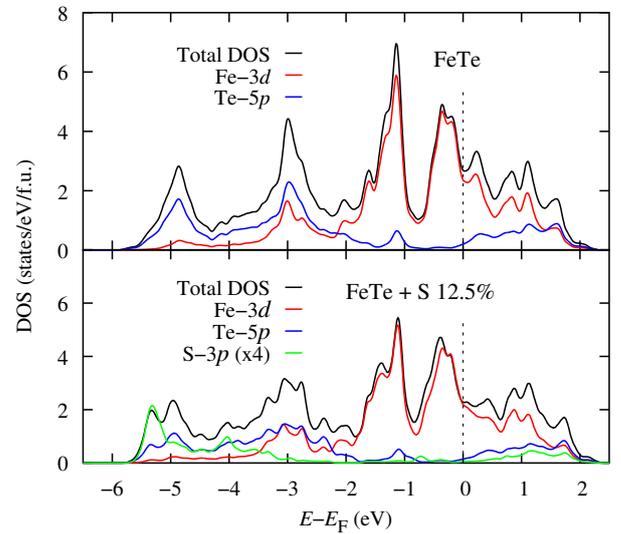

Figure 6: The non-magnetic total and partial densities of states of considered FeTe$_{1-x}$S$_x$ compounds with x = 0 and x = 0.125.

may turn the AFM2 into AFM1 state.

Interestingly, the AFM1 ground state was previously obtained for FeSe, Fe(Se,Te) and tensile strained FeTe [28, 37, 48, 49], similar to Fe pnictides [29, 44, 52]. It suggests that such ordering is a common feature of all iron superconductors. The results are consistent with the neutron scattering spectra as well, since spin excitations with $\mathbf{q} = (\pi, \pi)$ were observed in some superconducting Fe(Se,Te) solid solutions [30, 33, 34, 35].

The AFM2 to AFM1 phase transition is strongly related to the crystal structure [36, 37, 49]. The role of the S substitution is a compression of both lattice parameters and the average chalcogen position. These conditions can be achieved also by a partial Se substitution into Te sites or by *ab*-plane tensile stress applied to a thin film on an appropriate substrate. It was noticed that superconductivity with the highest $T_c$ of 15 K in FeTe$_{1-x}$Se$_x$ series occurs at x = 0.5, which yields an aver-age anion size of about 2.095 Å, while in FeTe substituted by a smaller S ion the superconductivity with $T_c^{max} = 10$ K occurs for FeTe$_{0.85}$S$_{0.15}$ with a very similar average anion radius ~ 2.1545 Å [14]. Thus, there exist the optimal distances $<h>$ in the system, which promote both the AFM1 spin correlations and the appearance of superconducting state [5, 6]. As indicated above, the structural considerations are of great importance in predictions of the magnetism and superconductivity of iron chalcogenides.

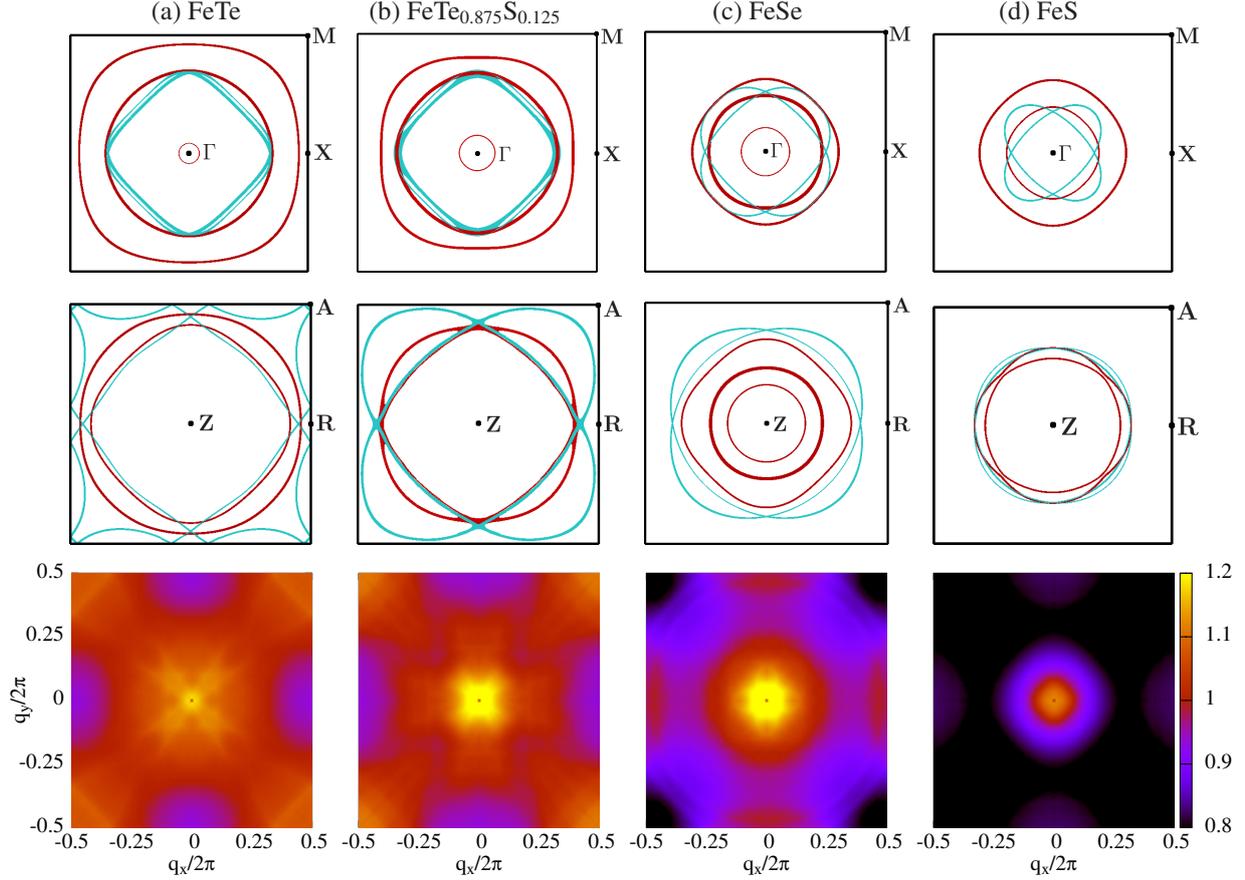

Figure 7: The Fermi surfaces of (a) FeTe and (b) FeTe$_{0.875}$S$_{0.125}$ in $k_z = 0$ (top) and $k_z = \pi$ (middle) plane of the folded Brillouin zone of the $2 \times 2 \times 1$ supercell obtained within non-spin-polarized calculations. The Fermi surface of (c) superconducting FeSe and (d) non-superconducting FeS are also drawn for comparison. The hole Fermi surface sheets are marked by black (red in the colour version) curves, while the the electron ones are marked by grey (cyan) curves [56]. The bottom panels show the real part of the static spin susceptibilities $\chi(\mathbf{q}, \omega \to 0)$ normalized by $\chi(\mathbf{q} = 0)$ for considered systems. The value of Re$\chi(\mathbf{q} = 0)$ for FeTe is about two times larger than those for other compositions (in arbitrary units).

### 3.2. Electronic structure of FeTe+S

To examine the non-magnetic electronic structure of Fe(Te,S), the bands and density of states (DOS) of FeTe and FeTe$_{0.875}$S$_{0.125}$ are plotted in Figures 5 and 6. These were obtained by the $2 \times 2 \times 1$ supercell calculations. As the substantial features of the electronic structure are sensitive to structural details [37, 38, 49, 53], the local positions $h_{Te}$ and $h_S$ are taken from the optimization of the magnetic phases.

The electronic structure of all iron chalcogenides consists of a few hole bands near the $\Gamma_0$ point and electron bands crossing the Fermi level ($E_F$) near the $M_0$ point of the Brillouin zone of the standard $1 \times 1 \times 1$ unit cell [25]. As expected, it is hard to compare the band structure of the $2 \times 2 \times 1$ supercell, presented in Fig. 5, with the ARPES spectra, because all these bands are folded into the $\Gamma$ point (more precisely, onto the $\Gamma$–Z line). It is clear, however, that the substitution with S atoms insignificantly changes bands and, hence, also Fermi surface and DOS in the vicinity of the Fermi level, since the S-$p$ states are located well below $E_F$. Meanwhile, a pseudogap near $E_F$ becomes more pronounced, similarly to that observed for superconducting Fe(Se,Te) systems [25]. It diminishes the DOS at $E_F$ from 2.69 states/eV/f.u. for pure FeTe to 2.28 states/eV/f.u. for the S-substituted system. For comparison, the DOS($E_F$) of superconducting FeSe is equal to 1.7 states/eV/f.u., while that of non-superconducting FeS amounts to 0.93 states/eV/f.u. for the optimized chalcogen position.

In Figure 7, the Fermi surfaces (for planes defined by $k_z = 0$ and $\pi$) and the real part of the non-interacting Lindhard spin susceptibilities $\chi$ (for $q_z = 0$) for FeTe and FeTe$_{0.875}$S$_{0.125}$ as well as for FeSe and FeS are visualized. The hole and electron Fermi surface sheets are centered at the $\Gamma$ and Z points of the folded Brillouin zone (FBZ) for appropriate $k_z$ planes and their overlapping provides the condition for ideal nesting [51]. The origin of these sheets can be recognized by a careful analysis of the band structures obtained for $2 \times 2 \times 1$ supercells (Fig. 5) or, alternatively, by comparison with the standard $1 \times 1 \times 1$ results and their folding where possible (not shown). The hole Fermi surface sheets, which come from the conventional zone center $\Gamma_0$, generally form more two-dimensional cylinders. The electron sheets derived from the original zone corner $M_0$ are, in turn, more three-dimensional and create ellipses rather than ideal circles in each $k_z$ plane (compare also with 1 Fe/cell results, *e.g.* [52]), with eccentricity increasing starting from $k_z = 0$ to $k_z = \pi$ [44].

The spin susceptibility is a measure of spin fluctuations derived from the band structure in the considered systems [25, 38, 51, 52]. Two types of such fluctuations have been determined for Fe chalcogenides within the $1 \times 1 \times 1$ unit cell. The dominant ones are connected with the tendency to the nesting-driven AFM1 ordering indicated by the maximum of the $\chi$ function at the $M_0(\pi, \pi)$ point. The others are connected with the tendency to FM ordering, revealed by the $\chi$ maximum at the $\Gamma_0(0, 0)$ point. In our approach, the two types of maxima associated with fluctuations are folded to the single $\Gamma(0, 0)$ point.

In the series of panels presented in Figure 7, the Fermi surface sections shrink with decreasing DOS at $E_F$. It was suggested that the intensity of the $(\pi, \pi)$ nesting, and thus corresponding AFM1 fluctuations, is the highest for FeTe among the considered iron chalcogenides because of the size of the Fermi surface [25]. However, the nesting for FeTe is strongly reduced by differences in both the shape and size of the hole and electron sheets, displayed in Fig. 7a. The hole Fermi surface sheets create nearly two-dimensional cylinders along the $k_z$ direction. In turn, the electron sheets in the ZRA plane ($k_z = \pi$) are so large that the part of their folded cylinders appears near the RA line in the FBZ boundaries. The overlapping of the hole and electron sheets occurs only near the $\Gamma$XM plane ($k_z = 0$). As a consequence, the spin susceptibility for FeTe does not exhibit sharp peak at the $(0, 0)$ point. A high value of DOS at $E_F$ (and thus $\chi(\mathbf{q} = 0)$) provides conditions for the appearance of the Stoner instability and suggests a tendency of FeTe to ferromagnetic ordering. Indeed, the phase transition of FeTe from the AFM to FM ordering was determined under hydrostatic pressure and was predicted to occur also under *ab*-plane compressive strain [11, 36, 37]. However, the high value of DOS at $E_F$ itself does not guarantee a strong condition for nesting.

For FeTe$_{0.875}$S$_{0.125}$, the Fermi surface consists of smaller hole and electron sheets (Fig. 7b) which, however, are better nested in the $\Gamma$XM plane, and the nesting is not so strong destroyed when going along $k_z$ as in pure FeTe. The response of FeTe$_{0.875}$S$_{0.125}$ to such a band structure, although being similar to that in FeTe, yields a strong resonance of the susceptibility $\chi$ at the $(0, 0)$ point. Such a resonance is also clearly resolved for superconducting FeSe and non-superconducting FeS. In the former system, it is caused by large both overlapping and parallel regions of hole and electron Fermi surface sheets, which provides the optimal condition for nesting and the appearance of enhanced spin fluctuations of the AFM1-type (Fig. 7c). If we compare $\chi(\mathbf{q})$ of FeSe for the $1 \times 1 \times 1$ unit cell (with peak of strong intensity at the $M_0$ point only) and for the $2 \times 2 \times 1$ supercells, we are sure that the peak is induced mainly by transitions from the holelike to electronlike sheets. Meanwhile, the DOS of FeS at the Fermi level is the lowest and the Fermi surface sheets are the smallest among considered Fe chalcogenides (Fig. 7d). The peak at $(0, 0)$ is quite well resolved, however, spin fluctuations are generally very tiny inducing neither magnetic order nor superconducting pairing.

Thus, nesting-driven spin fluctuations obtained from the NM state are well determined only for superconducting iron chalcogenides, which is consistent with our spin-polarized calculations. The findings indicate a great effect of the electron band structure on magnetic and superconducting properties of the Fe-based chalcogenides.

## 4. Conclusions

The magnetic phase transition from the double-stripe to single-stripe antiferromagnetic order was obtained for S-substituted FeTe by first principles calculations, corresponding to sulfur concentrations, for which the superconducting state was detected in the experiment [12, 13, 14, 15, 16]. The S substitution leads to a decrease of an average distance between chalcogen and iron atoms, in a similar way as under Se substitution or tensile stress in thin films of FeTe. It indicates that this transition between two antiferromagnetic states is related to the variation of the chalcogen atomic position.

The electronic structures of both FeTe and FeTe$_{0.875}$S$_{0.125}$ are characterized by the similar Fermi surfaces, however, nesting-driven spin fluctuations are well resolved only for the S-substituted compound. Thus, calculations curried out both with and without spin-polarization indicate the presence of spin fluctuations of the single-stripe type in the probed superconducting systems. The S substitution causes a suppression of the double-stripe antiferromagnetic order, which gives rise to the AFM1 correlations. The results point out that specific magnetic ordering plays an essential role in superconductivity of the iron chalcogenide.

## Acknowledgments

This work was supported by the National Science Center of Poland based on decision No. 2012/05/B/ST3/03095 and No. 2013/08/M/ST3/00927. Calculations were performed on ICM supercomputers of Warsaw University (Grant No. G46-13) and in Wrocław Center for Networking and Supercomputing (Project No. 158).